# A Limits Study of Memory-side Tiering Telemetry

Vinicius Petrucci
Micron Technology
San Jose, CA, USA
vtavarespetr@micron.com

Felippe Zacarias
Micron Technology
Austin, TX, USA
fvieirazacar@micron.com

David A. Roberts
Micron Technology
Boise, ID, USA
droberts@micron.com

*Abstract* — Increasing workload demands and emerging technologies necessitate the use of various memory and storage tiers in computing systems. This paper presents results from a CXL-based Experimental Memory Request Logger that reveals precise memory access patterns at runtime without interfering with the running workloads. We use it for software emulation of future memory telemetry hardware. By combining reactive placement based on data address monitoring, proactive data movement, and compiler hints, a Hotness Monitoring Unit (HMU) within memory modules can greatly improve memory tiering solutions. Analysis of page placement using profiled access counts on a Deep Learning Recommendation Model (DLRM) indicates a potential 1.94x speedup over Linux NUMA balancing tiering, and only a 3% slowdown compared to Host-DRAM allocation while offloading over 90% of pages to CXL memory. The study underscores the limitations of existing tiering strategies in terms of coverage and accuracy, and makes a strong case for programmable, device-level telemetry as a scalable and efficient solution for future memory systems.

## I. Introduction

Efficient memory tiering is crucial for optimizing hybrid memory/storage systems such as Host-local DRAM + memory expansion DRAM e.g. CXL, performance HBM + capacity LPDDR DRAM, DRAM+SSD, and future emerging memory technologies. Runtime systems must identify dynamically changing memory demands for different software objects and keep them in the optimal tier to meet performance and energy goals. Disparities in latency, bandwidth, and capacity between tiers create bottlenecks without careful management.

**Problem Statement.** While hot page detection is acknowledged as essential, current approaches—primarily instantiated at the software/kernel/host level—often suffer from limited coverage and accuracy, resulting in missing or misleading page promotion decisions and suboptimal workload performance. The collected performance data in this work reveals that the memory access patterns observed by the Operating System through standard profiling tools may not accurately reflect the true behavior of workloads.

**Related Work.** In recent years, several methodologies have emerged to enhance memory tiering and telemetry, addressing the shortcomings of traditional approaches. Heat-map telemetry units [2, 3] approximate the historical access frequency of a large address space using compact data structures. HOPP [4] augments the memory controller with a reverse page table to extract a sequence of virtual page addresses reflecting the real workload address locality. Filters then identify hot pages that are logged in a region of DRAM for later processing by Operating System page prefetch algorithms.

The **goal of this study** is to carry out a profile-driven analysis to explore the performance potential of memory-side tiering systems, offering understanding into their constraints and limits under representative workloads, while also benchmarking against state-of-the-art tiering strategies to contextualize the performance findings and insights. This work describes a flexible hardware-software collaborative tiering approach for memory-side tiering telemetry. It highlights the shortcomings of current tiering strategies in both coverage and accuracy, while advocating for programmable, device-level telemetry as a scalable and efficient approach for next-generation memory systems.

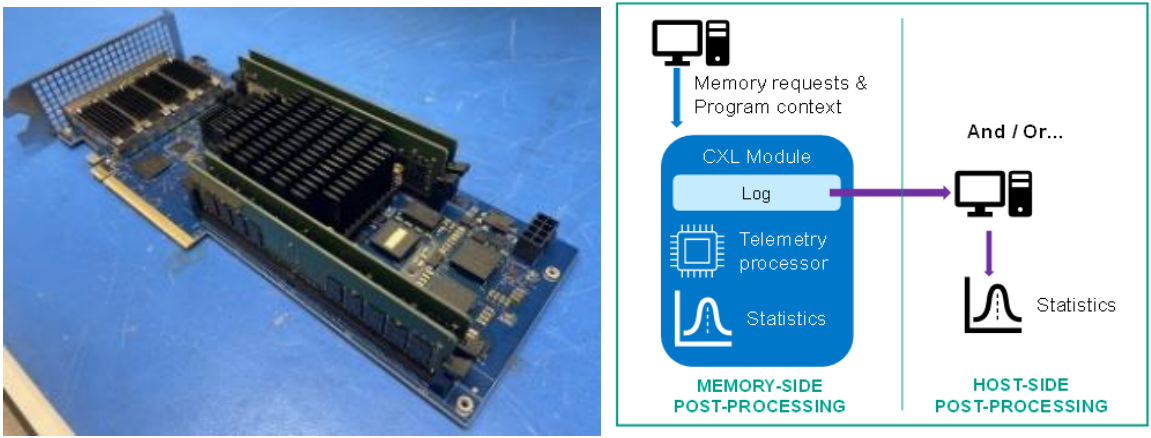

Figure 1. Micron's Experimental CXL-Based Memory Logger. It consists of an FPGA CXL DRAM card with 512 GB DDR4, where 256 GB is used for Main Memory (user data) and 256 GB for "CXL.mem" request packet log.

## II. CXL-BASED DRAM MODULE WITH DATA LOGGER

As shown in Figure 1, an FPGA-based CXL Type-3 module (DRAM memory expander) was implemented and connected to an Intel Emerald Rapids server. The module has 256 GB of DRAM for user-visible memory (instantiated as a CPU-less NUMA node in Linux) and another 256 GB DRAM for a CXL request log (used as a "devdax" memory type in Linux).

A Data Logger unit within the FPGA snoops and filters all CXL request packets. Depending on configuration settings, a subset of command types and fields are recorded with a timestamp as variable-length records in the log DRAM. The log can be read at any time by software, which can either save it to disk or process the trace immediately to evaluate user-defined tiering algorithms. Our main focus here is on the physical data addresses recorded in the log.

## III. STRATEGIES FOR PAGE PROMOTION

Page promotion strategies can be broadly classified into **OS-level**, **CPU-assisted**, and **memory-assisted approaches**. At the OS level, Linux's NUMA balancing (NB) memory tiering mode uses purely software-based heuristics, leveraging hints from minor page faults to promote pages to faster memory. CPU-assisted techniques, on the other hand, employ tools like Intel's PEBS (Precise Event-Based Sampling) to sample memory access, as demonstrated in a research study conducted by Google [1].

Memory-assisted approaches incorporate a **Hotness Monitoring Unit (HMU)** within memory modules, which can directly track all relevant hot pages within each monitoring range at the hardware level. This hardware-software collaboration holds the potential to enhance the accuracy of page promotion decisions and optimize the overall workload performance.

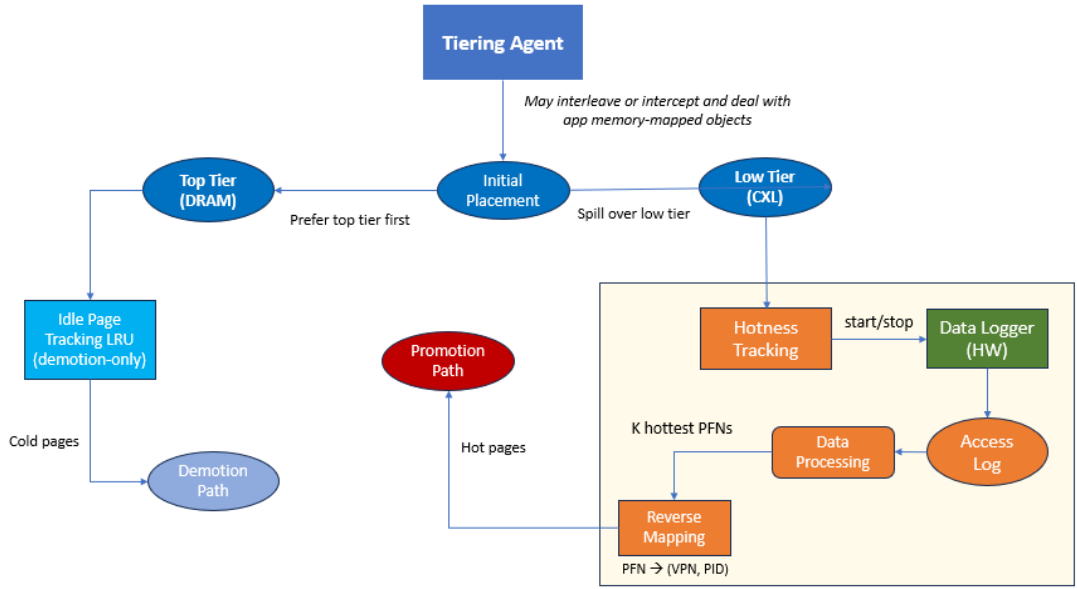

*Figure 2. Methodology for Evaluating Memory-side Tiering.*

Figure 2 illustrates the methodology developed to enable end-to-end performance evaluation of a memory tiering solution leveraging memory-side hotness tracking. It begins with a **Tiering Agent** that manages memory allocations by directing them to CXL memory or by binding application memory-mapped objects to specific memory tiers such as local DRAM or CXL modules. This agent can interleave or intercept allocation requests, ensuring that workloads are appropriately distributed across the available tiers.

The framework described does not implement page demotion itself, instead relying on the Linux kernel's existing LRU-based mechanisms for managing less frequently accessed pages. In contrast, page promotion is a central function within the memory tiering process: it selectively moves hot pages (accessed most frequently) from the low tier (CXL memory) to the top tier (Host DRAM). This ensures that critical data resides faster memory, optimizing system latency and overall performance.

With HMU, the telemetry system initiates the recording of access logs and applies reverse mapping to pinpoint pages within a workload's virtual memory space for the operating system's management. With CPU-assisted PEBS, hotness tracking simply counts virtual pages as the CPU records them. Hotness tracking continuously monitors memory access patterns, enabling the framework to identify which pages would benefit most from promotion. Data processing steps further analyze these logs, ensuring precise tracking and informed decision-making. Ultimately, hot pages are migrated using OS features from the low tier to DRAM, resulting in tangible reductions in memory latency and notable performance improvements for demanding workloads.

## IV. MEMORY TIERING WORKLOAD ANALYSIS

### A. Microbenchmark: mmap-bench

The mmap-bench microbenchmark allocates 10 GiB of memory, with 1 GiB being accessed for 90% of the execution. Within this frequently accessed region, the precise number of pages eligible for promotion is calculated as K = 262,144 (4KiB) pages. During the profiling phase, memory allocation operations are directed to CXL memory, allowing for device-side monitoring of page access patterns. For a fair comparison with PEBS-based tiering, both the HMU and PEBS data collectors are started simultaneously to generate a list of tuples containing data addresses and their corresponding access counts. Following the warm-up period, the top-K pages are selected for promotion to higher-performance DRAM, ultimately improving system memory access latency and workload performance.

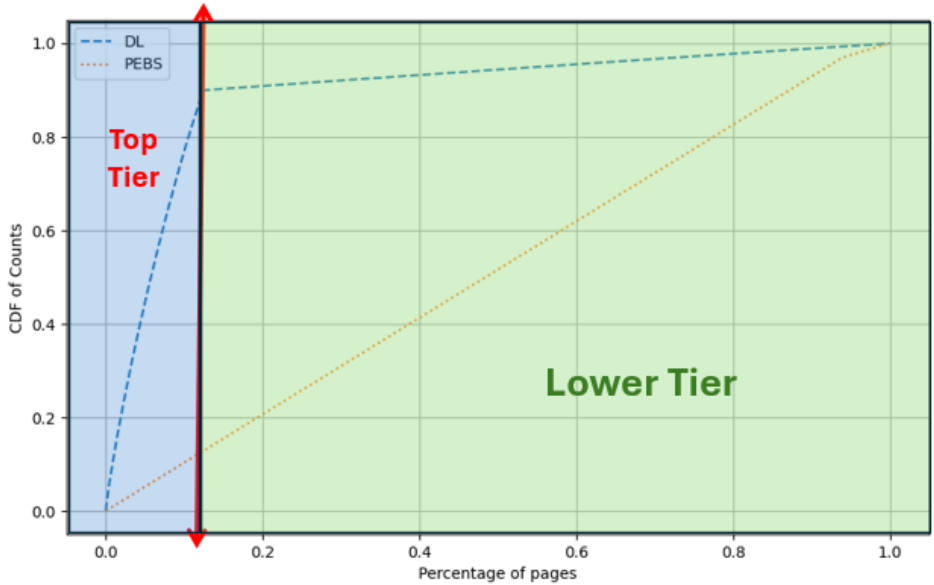

*Figure 3. Hotness Distribution: mmap-bench.*

As shown in Figure 3, the HMU (labeled Data Logger or DL) precisely captures workload behavior, revealing that ~ 10% of pages account for ~ 90% of memory accesses. PEBS sampling fails to accurately reflect workload hotness, leading to an assumption of evenly distributed memory access. The analysis covers only accessed pages.

The **"Oracle" Hotness-based Tiering** approach, as illustrated in Figure 3, involves promoting workload pages based on hot-page lists generated by HMU. Thus, the usable capacity of the top tier is effectively limited to K pages (size of the workload hot region), which represents the maximum number of hot pages that can be migrated from lower tiers. This analysis examines the accuracy of hot-page detection by PEBS versus HMU and their effects on workload performance (reads/second). While PEBS achieved 87% accuracy confirmed by HMU, it only promoted 6% of K pages as hot. HMU-based promotions delivered much better results, with performance **2.94 times faster than PEBS-based tiering**.

A second analysis compared NB (Linux) page migration for the initial K hot pages against HMU, focusing on hot-page identification accuracy and workload performance. Results showed 75% overlap between NB and HMU selections, indicating NB had 25% lower promotion accuracy. For workload performance, **HMU was 1.73x faster than NB**, highlighting the benefits of device-level hotness tracking for memory tiering.

### B. Real Workload: Deep Learning Recommender Model

This section applies the previously described methodology ("Oracle" Hotness-based Tiering) to examine the function of CXL-attached memory as a secondary tier for storing extensive user embedding tables within a Deep Learning Recommender Model (DLRM). Performance is evaluated using the FBGEMM (using production-level datasets from Meta[1]) split table benchmark, which quantifies the throughput and latency of batched embedding bag operations, which are the core computational kernels in large-scale personalized recommendation systems.

Meta's dataset demonstrates the sparse nature of access patterns within large-scale embedding tables. In a typical table, for example, the embedding model comprises 5.12 billion parameters, occupying 20.48 GB of memory. However, only 2.95 GB of embedding weights are accessed per batch, indicating that only 14% of the total parameters are utilized during a typical inference pass. This highlights a significant opportunity for memory optimization, such as tiered memory systems, to reduce the footprint of inactive parameters without impacting model performance.

---

[1] https://github.com/facebookresearch/dlrm_datasets/tree/main

*Table 1. DLRM inference comparison.*

| Method | Average Inference Time (us) | Pages Promoted | Speed Comparison | Top Tier Footprint (GB) |
|---|---|---|---|---|
| **HMU** | **65,454** | **486,587** | **1.94x faster than NB** | **1.85** |
| NB | 127,294 | 481,683 | - | 1.92 |
| DRAM-only | 63,324 | - | 1.03 faster than HMU | 20.48 |

Table 1 summarizes the DLRM inference results, showing that **HMU performs 1.94 times faster than the existing NB solution**. For fairness, NB had two iterations to promote hot candidates like HMU. Compared to the ideal DRAM-only baseline, HMU is only 3% slower and uses just 9% of the DRAM footprint for the hottest pages, consisting of 1.85 GB out of 20.48 GB required for embedding tables.

## V. Conclusions

This study highlights the following key takeaways regarding memory tiering strategies. **PEBS lacks coverage** due to its inherent sampling methodology, which fails to capture all pertinent hot pages, resulting in the omission of certain pages that should be promoted to the fast memory tier. While increasing coverage is possible, it would come at the cost of higher CPU overhead.

On the other hand, **Linux NB lacks accuracy** as it may misclassify true "super" hot pages, leading to inefficient use of fast memory by promoting pages that do not necessarily benefit from it. In contrast, **device-level hotness tracking** can meet both coverage and accuracy requirements without incurring any CPU/host overhead, making it a superior solution for optimizing memory tiering.

## VI. Opportunities & Future Work

Due to the variety of workloads, telemetry, and algorithms for managing data between multiple tiers, a flexible solution may be effective. Programmable memory-side telemetry, driven by near-memory computing (NMC), counters and logging, could enable automatic reactive placement with HMU accuracy, proactive data movement such as trace-based intelligent prefetching, and programmer/compiler hints to rank data based on prior knowledge. Key research areas include reducing DRAM needed for logging and optimizing telemetry processors for high request rates while maintaining programmability. Using Host processors for hardware memory request log analysis can be improved with NMC to generate statistics more efficiently and save I/O bus bandwidth.

Future work will align with the Linux kernel community to enhance memory tiering under the CXL 3.2 specification, which introduces CHMU for hotness tracking in composable memory systems. The OCP (Open Compute Project) community has previously suggested hotness tracking [5]. The aim is to standardize such a hotness monitoring capability across vendors and platforms.